\documentclass[aps,pre,showpacs,
amssymb,twocolumn,groupedaddress]{revtex4}
\usepackage{graphicx}
\usepackage{amsmath}
\usepackage{amsfonts}
\usepackage{amssymb}
\usepackage{epsfig,libertine}
\usepackage[libertine]{newtxmath}
\usepackage{color}
\usepackage{dsfont, hyperref, xcolor}
\usepackage{comment}
\usepackage{enumerate}
\newcommand{\Tr}[1]{\text{Tr}\left\{#1\right\}}

\newcommand{\bra}[1]{\langle#1\vert}
\newcommand{\ket}[1]{\vert#1\rangle}
\newcommand\braket[2]{\langle#1|#2\rangle}

\begin{document}

\title{Hilbert space fragmentation in a long-range system}

\author{Gianluca Francica, Luca Dell'Anna}
\address{Dipartimento di Fisica e Astronomia e Sezione INFN, Universit\`{a} degli Studi di Padova, via Marzolo 8, 35131 Padova, Italy}

\date{\today}

\begin{abstract}

We study the role of the interaction range on the Hilbert-space fragmentation and many-body scar states
considering a spin-1/2 many-body Hamiltonian describing a generalized Fredkin spin chain.
We show that both scar states and weak fragmentation of the Hilbert space survive for almost any range of the coupling.
{\color{black}Moreover, when the interaction range is small enough, there are sectors with definite symmetries such that the ratio between the dimension of their largest fragment and their dimension decays algebraically with the system size}. 
Finally we investigate the effects of such structures of the Hilbert space on the out-of-equilibrium dynamics, triggered by certain initial states, characterized by either local persistent oscillations or non-uniform stationary profile of the magnetization.

\end{abstract}

\maketitle

\section{Introduction}
A closed system given in a certain initial state can thermalize, i.e., the expectation values of local observables at long times tend to the expectation values calculated with respect to a thermal state which does not depend on the details of the initial state, except for the energy.

Thermalization can be explained by using the eigenstate thermalization hypothesis (ETH)~\cite{Deutsch91,Srednicki94}.
In simple terms, an energy eigenstate satisfies the ETH 
if the reduced state of a small part of the system is a thermal state. Thus, the matrix element of any local
operator expressed in a basis of energy eigenstates
is a smooth function of the energy, and, if diagonal, is determined by its average value in an appropriate Gibbs density matrix~\cite{Rigol08,Polkovnikov11,DAlessio16}. There are cases where
ETH is violated (see, e.g., the reviews Refs.~\cite{Papic21,Moudgalya22}). In general, ETH is said to be strongly satisfied if every eigenstate satisfies ETH, and weakly satisfied if almost all eigenstates satisfy ETH except a zero measure set of them. Roughly speaking, in both cases a randomly chosen initial state with a narrow energy distribution will thermalize after long times, however if ETH is only satisfied weakly we can always find some initial states having narrow energy distributions which do not thermalize.

We expect that for a non-integrable system without any symmetries the time evolution of a random initial state generates the full Hilbert space and ETH is strongly satisfied. We recall that integrable systems have a level statistics which follows a Poisson distribution, whereas non-integrable systems have a level statistics which can be described by a random matrix theory ensemble~\cite{DAlessio16}. However, there are some exceptions, i.e., cases where ETH is not satisfied, for instance, some integrable systems having an extensive number of conserved quantities, which, in one spatial dimension, can be exactly solved by a Bethe ansatz procedure. In this case the long time expectation values are given by a so-called generalized Gibbs ensemble~\cite{Essler16}. There are also some
 interacting systems exhibiting many-body localization, in the presence of strong disorder or quasiperiodicity, where again an extensive number of conserved quantities can be constructed (see, e.g., the reviews Refs.~\cite{Nandkishore15,Abanin19}). In these cases both strong and weak ETH are violated.

In a chaotic system, ETH can be only weakly satisfied in the presence of quantum many-body states called scars states (see, e.g., Ref.~\cite{Serbyn21} for a recent review), which form a zero measure set of highly excited eigenstates that do not satisfied ETH and thus 'scar' the spectrum of ETH-satisfying eigenstates.
Furthermore, strong ETH is not satisfied if there is fragmentation of the Hilbert space~\cite{Sala20,Khemani20}, namely there is a number of dynamically disconnected invariant subspaces, called Krylov subspaces, which exponentially grows with the size of the system. We can have weak (strong) fragmentation if there is (is not) a dominant subspace and weak ETH is (is not) satisfied.
Moreover, Krylov-restricted ETH can be satisfied leading to surprising behavior, e.g., thermal expectation values of observables which are spatially non-uniform~\cite{Moudgalya21}.
{\color{black}Fragmentation is known to appear in dipole-moment or center-of-mass conserving systems~\cite{Sala20,Khemani20,Moudgalya21,Pai19,Rakovszky20,Morningstar20}, in other systems like the t-V model~\cite{DeTommasi19}, certain one-dimensional models with strict confinement~\cite{Yang20,Chen21,Bastianello22}, PXP models~\cite{Mukherjee21,Mukherjee221}, in the presence of frustration~\cite{Lee21,Hahn21} and in polar lattice gases~\cite{Li21}. Furthermore, spin chains such as the Fredkin~\cite{Langlett21}, Motzkin~\cite{Richter22}, Pair-Flip~\cite{Moudgalya222}, folded XXZ~\cite{Zadnik21}, and
Temperley-Lieb~\cite{Read07,Moudgalya222} spin chains, can exhibit fragmentation. 
Although there are several systems that show fragmentation only in one dimension~\cite{Moudgalya222}, 
it can occur also in systems with dimension larger than one~\cite{Khudorozhkov22,Chattopadhyay23}. }
{\color{black} Moreover, many-body scars resulting from Hilbert space fragmentation have been explored in Refs.~\cite{scar1,scar2,scar3,scar4,scar5}.}

To explain how ETH implies thermalization, we consider a non-degenerate spectrum of a Hamiltonian $H$, i.e., $E_i\neq E_j$ whenever $i\neq j$, and an initial state $\ket{\psi_0}$ with a narrow energy distribution with mean value $E=\bra{\psi_0}H\ket{\psi_0}$, so that $\sum_{i: E_i\approx E} p_i \approx 1$, where $p_i=|\braket{E_i}{\psi_0}|^2$ and $\ket{E_i}$ is the eigenstate of $H$ with eigenvalue $E_i$. The time-evolved state is $\ket{\psi_t}=e^{-iH t}\ket{\psi_0}$, and the time-averaged density matrix is
\begin{equation}
\bar{\rho} = \lim_{\tau\to \infty}\frac{1}{\tau} \int_0^\tau \ket{\psi_t}\bra{\psi_t} = \sum_i p_i \ket{E_i} \bra{E_i}\,,
\end{equation}
so that it is equal to the so-called diagonal ensemble. {\color{black} If there is thermalization the long-time expectation value of an observable is determined by $\bar{\rho}$.}
Since the initial state has a narrow energy distribution, we get
\begin{equation}
\bar{\rho}\approx \sum_{i: E_i \approx E}p_i \ket{E_i} \bra{E_i}\,.
\end{equation}
In principle, this state can depend on the initial state details through the distribution $p_i$. Anyway, if the states $\ket{E_i}$ satisfy the ETH, they are thermal, and for $E_i\approx E$ the reduced states of a small part of the system are thermal states with inverse temperature $\beta_{E_i}\approx \beta_E$, and of course $\bar{\rho}$ locally is approximately equal to a thermal state with inverse temperature $\beta_E$.
More precisely, ETH postulates that the matrix elements of a local observable $O$ with respect to the basis of the energy eigenstates, $O_{mn} = \bra{E_m} O\ket{E_n}$, are
\begin{equation}\label{eq. ETH}
O_{mn} = O(\bar{E}) \delta_{m,n} + e^{-S(\bar{E})/2} f_O (\bar{E},\omega) R_{mn}
\end{equation}
where $O(\bar{E})$ is
a smooth function of the energy
$\bar{E}=(E_m+E_n)/2$,
corresponding to the expectation value with respect to the microcanonical ensemble with energy $\bar{E}$,
while $\omega=E_m-E_n$,  $S(\bar{E})$ is the equilibrium entropy of the system, so that the temperature is given by $\beta_E=\partial_E S(E)$, $f_O (\bar{E},\omega)$ is some observable-dependent function and $R_{mn}$ is some random variable with zero mean and unit variance.
We note that the entropy $S(E)$ is generally peaked at the middle of the spectrum,
so that for those energies the effective temperature is very large and
the average $O(\bar{E})$ is given by a local completely mixed state and $e^{-S(E)/2}\sim 1/\sqrt{\mathcal D}$, where $\mathcal D$ is the dimension of the Hilbert space (almost all states are contained near the entropy peak in the limit of large system size). Equation~\eqref{eq. ETH} is, then, equivalent to a random matrix theory description~\cite{DAlessio16}.

{\color{black} In this paper, we focus on a longer-range generalization of the Fredkin chain studied in Ref.~\cite{Langlett21}.
In principle, these kind of models can be realized experimentally by cold atoms in optical lattices, and a three-spin interaction can be designed as in Ref.~\cite{Pachos04}.
For the nearest neighbor case, strong fragmentation and scars can coexist, while what happens by increasing the interaction range was so far 
an open question. It was supposed that longer range interactions could be detrimental for fragmentation. Interestingly, we find that weak fragmentation and scar states survive for long-range interactions. }

\section{Model}
Let us now introduce the following Hamiltonian
\begin{equation}
H= \sum_{\ell=1}^{L-3} d_\ell \sum_{i=2}^{L-\ell-1} \left({\cal P}^\uparrow_{i-1} \otimes
{\cal S}_{i,i+\ell}
- {\cal S}_{i,i+\ell}
\otimes {\cal P}^\downarrow_{i+\ell+1}\right)\,,
\end{equation}
where we defined the projectors
\begin{eqnarray}
{\cal P}^\sigma_i &=& \ket{\sigma_i}\bra{\sigma_i},\\
{\cal S}_{i,j}&=&\ket{s_{i,j}}\bra{s_{i,j}}
\end{eqnarray}
with $\sigma=\uparrow,\downarrow$, and the singlet state
\begin{equation}
\ket{s_{i,j}}=\big(\ket{\hspace{-0.05cm}\uparrow_i \downarrow_j}-\ket{\hspace{-0.05cm}\downarrow_i\uparrow_j}\big)/\sqrt{2}.
\end{equation}
Differently from the Fredkin chain~\cite{DellAnna16,Salberger16}, the minus sign in the second term produces an interference in the Fredkin moves, and the sector of the Dyck paths is fragmented in many disconnected subspaces.
The nearest neighbor case, i.e., $d_{\ell>1}=0$, has been investigated in Ref.~\cite{Langlett21}, and exhibits Hilbert space fragmentation and scar states. Here, we aim to study how a longer range interaction affects those peculiar behaviors. Hence, we consider an interaction of range $r$, i.e., $d_\ell\neq 0$ for $\ell\leq r$ and $d_\ell = 0$ for $\ell> r$. Of course, $r\leq L-3$, and for $r=L-3$ we get the fully connected case.

Let us introduce the relevant symmetries.
The total magnetization $S_z = \sum_{i=1}^L \sigma^z_i$ is a symmetry for any range $r$, i.e., $[H,S_z]=0$, while the domain-wall number $n_{\textrm {dw}}=\sum_{i=1}^{L-1}\sigma^z_i \sigma^z_{i+1}$ does not commutate with the Hamiltonian for $r>2$, and it is a symmetry only for $r\leq 2$, where $\sigma^\alpha_i$ with $\alpha=x,y,z$ are the Pauli matrices on the site $i$. Furthermore, for any $r$, there are the two symmetries $S_1=\sigma^z_1$ and $S_L=\sigma^z_L$, which fix the spins at the boundaries. Of course, there is also the parity symmetry $P = \otimes_{i=1}^L \sigma^z_i$.
We note that there is also a chiral symmetry represented by the operator $C=\otimes_{i=1}^L \sigma^x_i$ that anticommutates with the Hamiltonian, so that the spectrum is symmetric with respect to zero energy.

\section{Hilbert space fragmentation}
\begin{figure}
[t!]
\centering
\includegraphics[width=0.89\columnwidth]{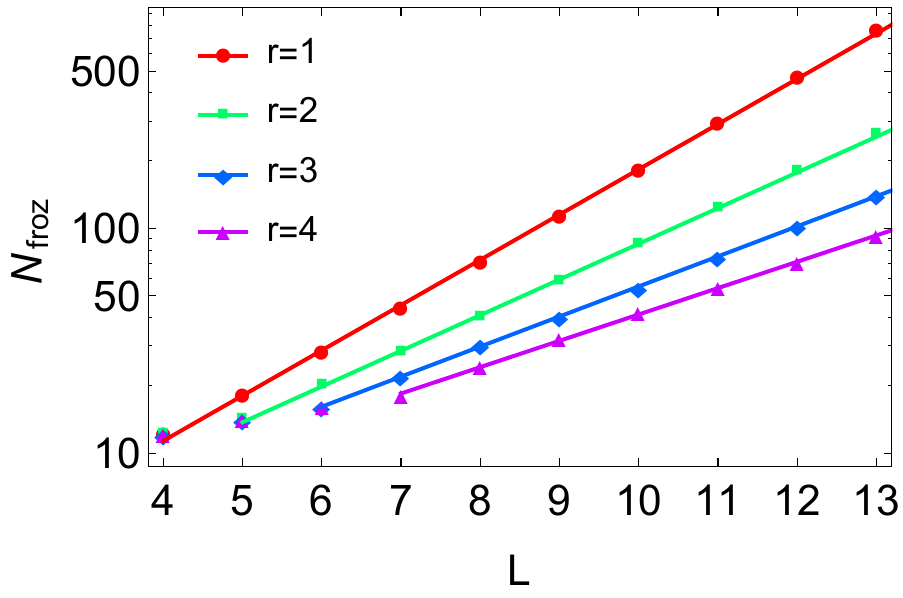}
\caption{$N_{froz}$ as a function of $L$ for different values of $r$. The solid lines are obtained by fitting the points, for $L\geq 3 +r$, with the function $\propto a^L$.
}
\label{fig: nfroz}
\end{figure}

The Hilbert space $\mathcal H$ can be decomposed into dynamically disconnected Krylov subspaces as
\begin{equation}
\mathcal H = \bigoplus_{n=1}^{K} \mathcal K_n\,,
\end{equation}
where the Krylov subspace $\mathcal K_n$ is spanned by $\ket{\psi_n}$, $H \ket{\psi_n}$, $H^2 \ket{\psi_n}$ and so on, for a certain so-called root state $\ket{\psi_n}$ which is a product state having defined symmetries. If there are only conventional symmetries, e.g., $S_z$, the number $K$ of Krylov subspaces grows at most polynomially upon increasing the system size $L$. In contrast, if there is fragmentation of the Hilbert space, then $K\sim e^{c L}$ in the thermodynamic limit $L\to\infty$. There are Krylov subspaces with dimension one as well as exponentially large subspaces (see Appendix~\ref{app. Krylov}).
The one-dimensional Krylov subspaces are called frozen states. For our model
the total number of frozen states is $N_{froz}=12+2k$ for a size $L=4+k$ with $k=0,1,\ldots,r-1$, whereas
{for $L>3+r$ an exponential growth starts, i.e., $N_{froz}\sim a_r^{L-r-3}$} 
as $L\to\infty$, where $a_r$ decreases upon increasing $r$ (see Fig.~\ref{fig: nfroz}).
Concerning the total number of Krylov subspaces, we get $K=30+6 (k-2)$, if $k\geq 2$, or $K=14+8 k$, if $k<2$, for a size $L=4+k$ with $k=0,1,\ldots,r-1$.

Let us calculate the total number of sectors with definite symmetries, $S_z$, $S_1$ and $S_L$, for $r\geq 3$. We note that, in this case, we get 4 distinct combinations of the boundaries spins, and, in the bulk, there are $L-1$ sectors with defined magnetization, then we have $4(L-1)$ sectors with defined symmetries.
As a result, for a size $L=4+k$ we get a number of sectors equal to $12+4 k$ which is always smaller than the total number of Krylov subspaces for any range $r$.
We deduce, therefore, that there is a fragmentation of the Hilbert space (with an exponential number of Krylov subspaces)
for any {range $r$ such that $r/L\to b$ as $L\to \infty$, with $b\in[0,1)$. 
Thus, there is almost always fragmentation, except when the range is $r= L-c_r(L)$, for any number $c_r(L)\geq 3$ such that $c_r(L)/L\to 0$, i.e., the range is asymptotically linear in the size with a linear coefficient equal to one. In this case, the number of Krylov subspaces will grow slower than exponential.}

In order to characterize the fragmentation of a sector of dimension $D$ defined by certain symmetries, we consider the largest Krylov subspace
with dimension $D_{max}$, belonging to this sector.
For instance, if the symmetries are $S_z$, $S_1$ and $S_L$, as in the case with $r\geq 3$, the dimension of the sector is given by
$D=\binom{L-2}{(L+S_z-S_1-S_L)/2-1}$. We can get, therefore, sectors with dimension $D$ which grows exponentially with $L$, e.g., for $S_z\sim 0$, or polynomially, e.g., linearly for $S_z=\pm(L-2)$ and $S_1=S_L=\pm 1$. If there is also the symmetry $n_{\textrm{dw}}$, as in the cases with $r\leq2$,
$D$ has a different expression. For example for $L$ even, namely $L=2+2k$, and $S_z=0$, $S_1=1$, $S_L=-1$, we get
$n_{\textrm{dw}}=L-1-2n_{k}$, where the kink number is $n_k=1+2 x$, with $x=0,1,2,\ldots,(L-2)/2$, and the dimension $D=\binom{k}{x}^2$.
Given $D$ and $D_{max}$, we have the so-called strong fragmentation or weak fragmentation if $D_{max}/D\to 0$ or $D_{max}/D\to 1$, respectively, in the thermodynamic limit \cite{Sala20}.

The case $r=1$ has been investigated in Ref. \cite{Langlett21} where it has been shown that there are sectors of dimension $D$ such that the dimension of their larger Krylov subspace $D_{max}$ grows exponentially, still the ratio $D_{max}/D\to 0$, vanishes exponentially for $L\to \infty$, revealing strong fragmentation.

For $r=2$, if there is fragmentation of a sector with dimension $D$, that sector is fragmented in frozen states plus at most one dominant larger Krylov subspace. In detail, the sector will be the direct sum of all the $N'_{froz}$ frozen states in that sector and the complementary subspace having dimension $D_c=D-N'_{froz}$.
One can check 
that, for any fragmented sector of dimension $D$, the quantity $D_c-D_{max}$ 
is zero (there is only one Krylov subspace with dimension larger than one, since $D_{max}\ge 1$ and if $D_c\neq 0$ then necessarily $D_c>1$, by definition) or -1 (all the Krylov subspaces are frozen states, namely $D_c=0$ and $D_{max}=1$).
Actually, when $D-D_{max}-N'_{froz}=0$ we find always only one frozen state in the sector, i.e., $N'_{froz}=1$, thus getting only two Krylov subspaces in the sector, 
and, therefore, we get weak fragmentation, since $D_{max}\simeq D$ for large $L$.
Furthermore, we can get a kind of strong fragmentation, i.e., $D_{max}/D\to 0$ as $L\to \infty$,  when $D=N'_{froz}>1$. 
This situation occurs, for instance, in the sectors with dimension $D=L-4$ (i.e., which grows linearly with $L$), and for $S_z=\pm(L-6)$, $n_{dw}=L-5$, $S_1=S_L=\pm 1$, which are the largest sectors among those with dimension $D=N'_{froz}>1$. In this case $D_{max}/D\sim 1/L\to 0$ as $L\to \infty$, i.e., the ratio does not tend to zero exponentially.
We note that this kind of non-exponential strong fragmentation is also 
present for $r=1$.

\begin{figure}[t!]
\centering
\includegraphics[width=0.89\columnwidth]{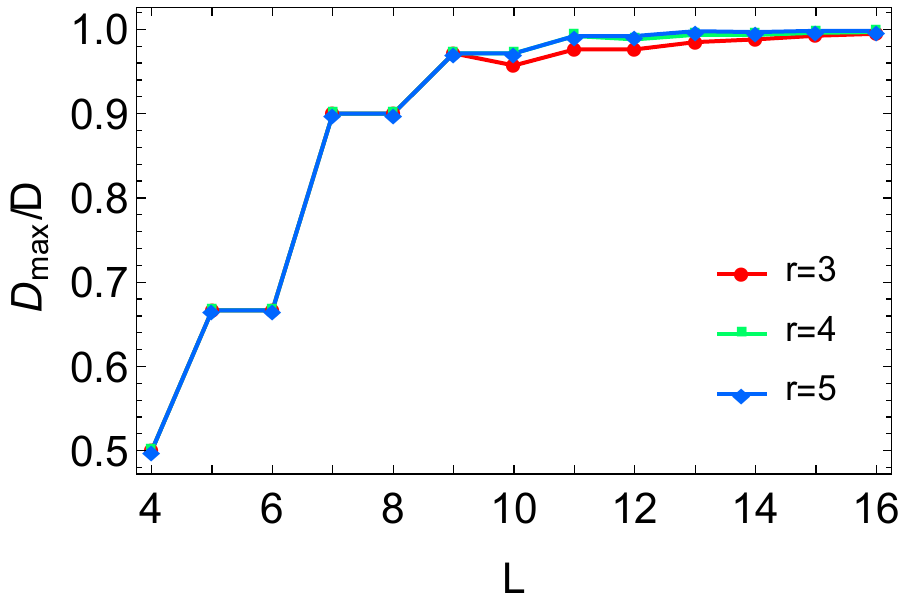}
\caption{$D_{max}/D$ as a the function of $L$ for different values of $r$. We consider the sector with $S_z=0$ or $1$ for $L$ even or odd, $S_1=1$ and $S_L=-1$.}
\label{fig: ratio}
\end{figure}

\begin{figure}
[t!]
\centering
\includegraphics[width=0.4\columnwidth]{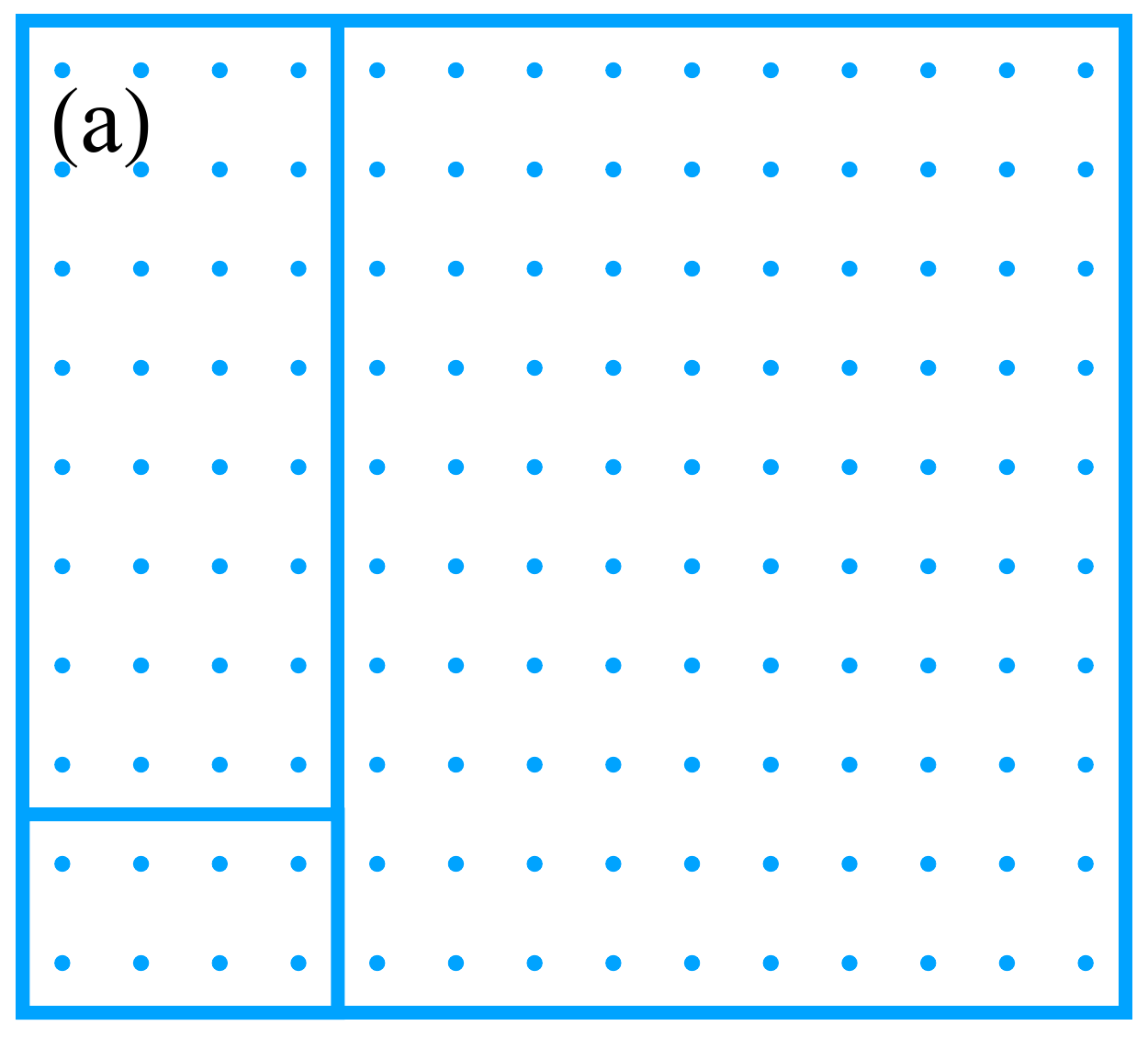}\includegraphics[width=0.4\columnwidth]{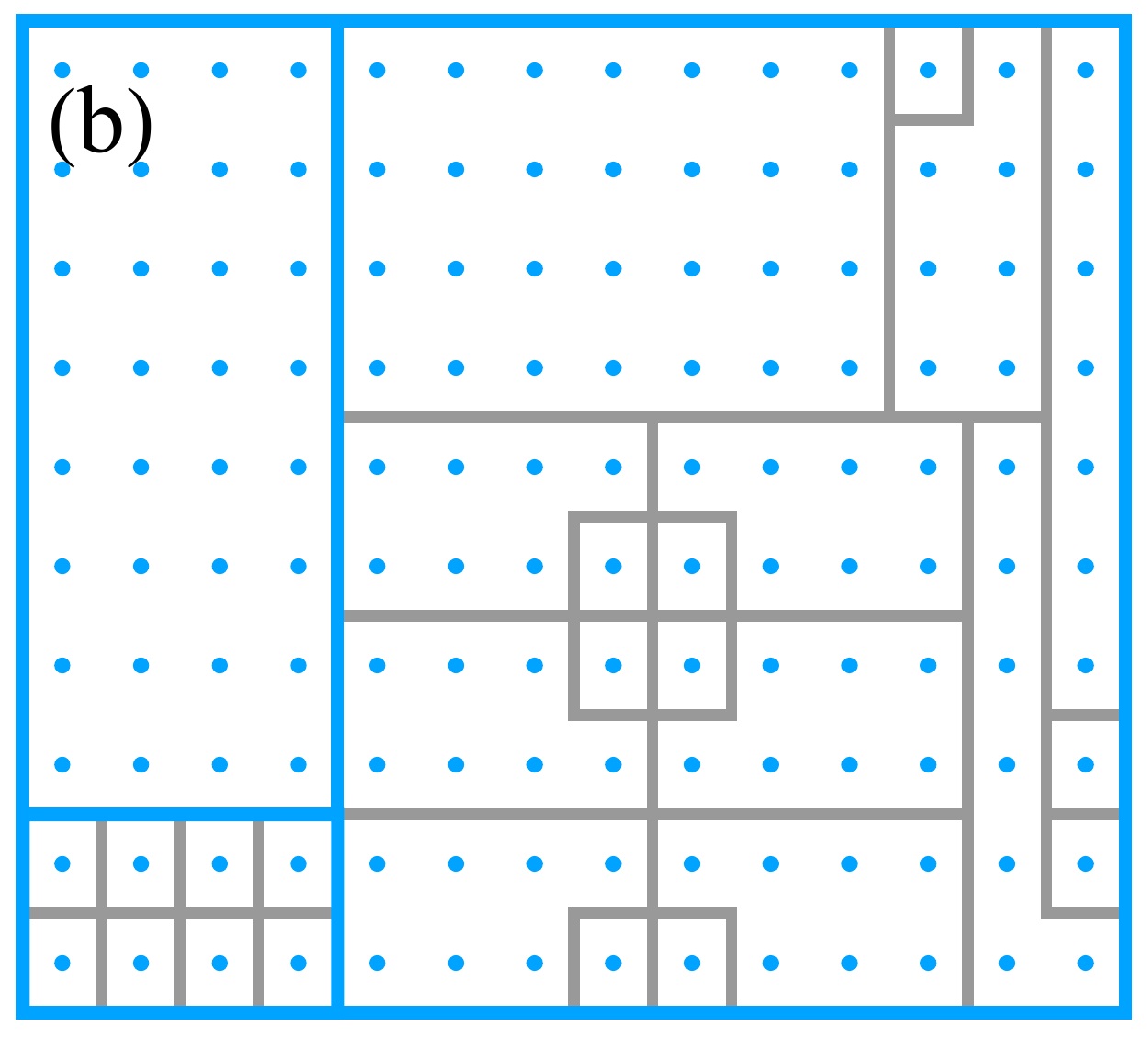}\\
\includegraphics[width=0.4\columnwidth]{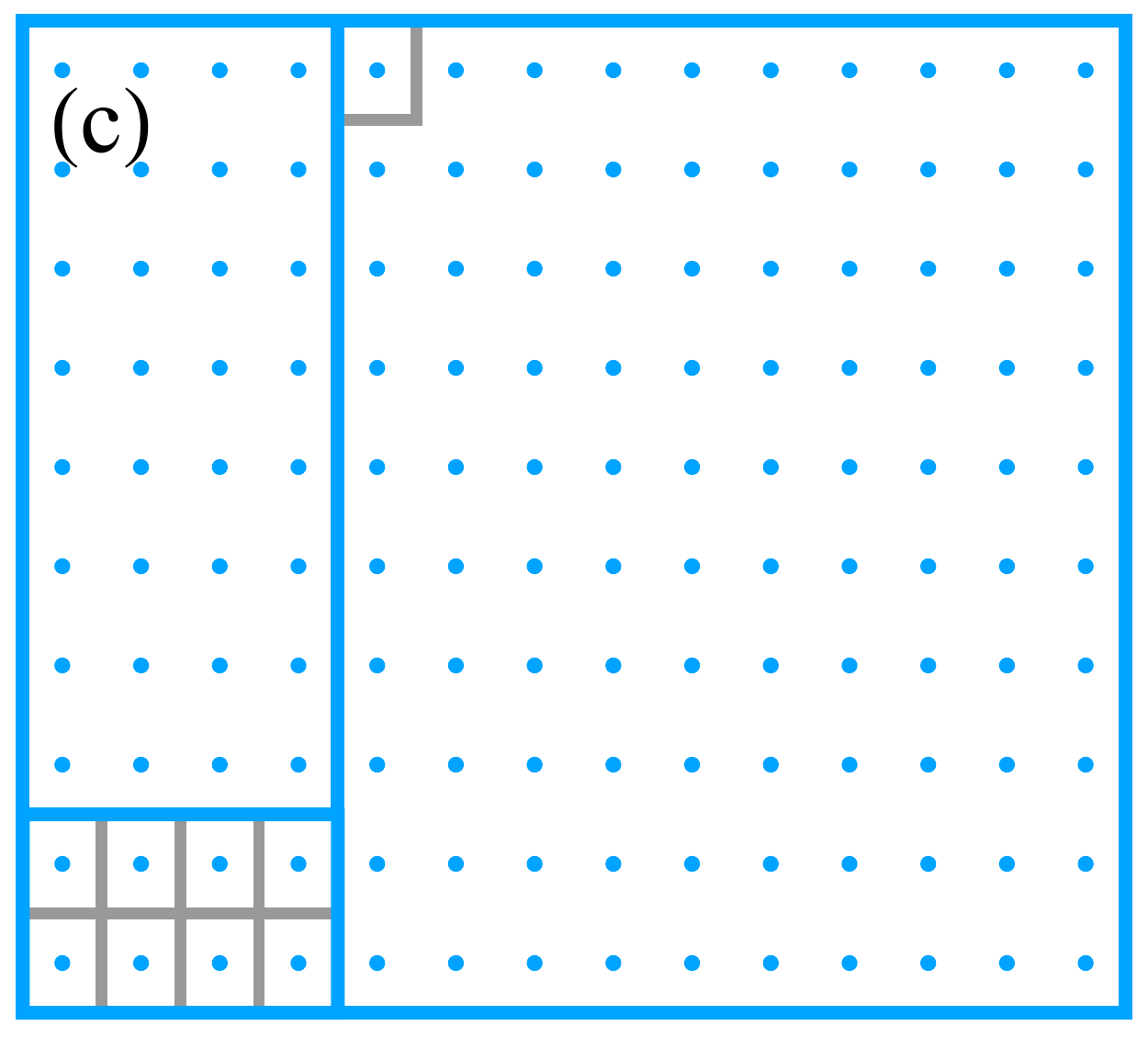}\includegraphics[width=0.4\columnwidth]{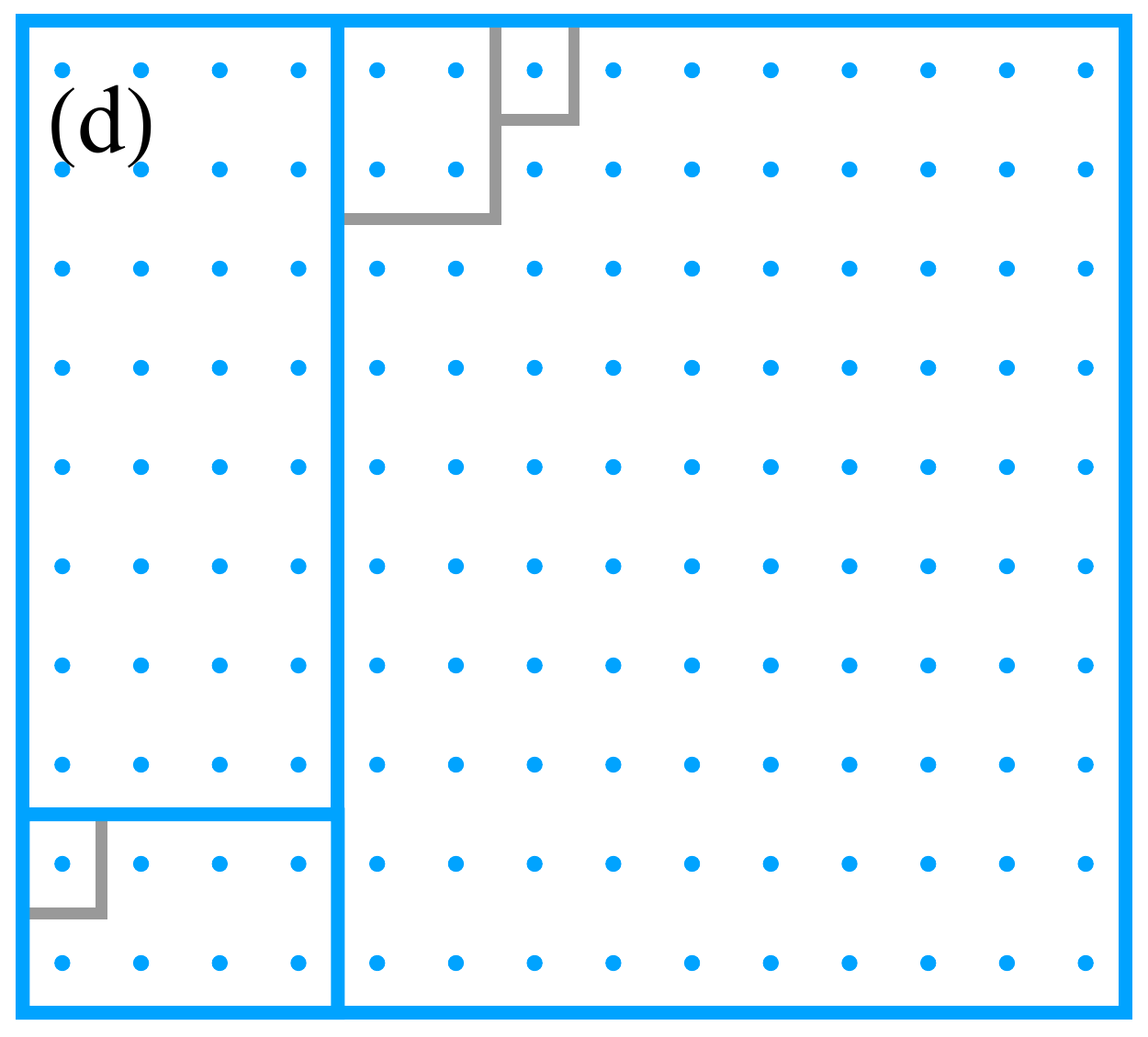}
\caption{A pictorial representation of the fragmentation. In Fig.~(a) we represent the sectors with definite symmetries as regions containing certain states represented by points
{(the largest sector with $100$ points is obtained for $L=12$, $S_z=0$, $n_{\textrm{dw}}=1$, $S_1=-S_L=1$ and $r\leq 2$. For $r>2$ the largest sectors have different symmetries although we pictorially represent them in the same way).}
The sectors will be fragmented in Krylov subspaces (regions with gray boundary lines in Figs.~(b,c,d)). For $r=1$ (see Fig.~(b)), there are sectors fragmented in frozen states, with dimensions linearly dependent on $L$, and sectors fragmented in an exponential number of Krylov subspaces, with the largest Krylov subspace with dimension $D_{max}$ which grows exponentially with $L$, so that $D_{max}/D\to 0$. For $r=2$ (see Fig.~(c)), there are still sectors fragmented in frozen states, with a dimension linearly dependent on $L$, and sectors fragmented in one frozen state and one dominant large Krylov subspace with a dimension $D_{max}$ such that $D_{max}/D\to 1$. For $r>2$ (see Fig.~(d)), there are fragmented sectors containing a dominant large Krylov subspace with a dimension $D_{max}$ such that $D_{max}/D\to 1$.
}
\label{fig: schema}
\end{figure}

In contrast, for $3\leq r\leq L-3$, we do not find strong fragmentation in any sector. In particular, we find that for the sectors such that $D-D_{max}-N'_{froz}=-1$ we get $N'_{froz}=1$, i.e., the sector is just a frozen state ($D=D_{max}=1$), therefore, trivially there is not fragmentation. However, for any $r\geq 3$, there are sectors of dimension $D$ with definite symmetries such that $D_{max}<D$ and $D_{max}/D \to 1$ as $L$ increases (see Fig~\ref{fig: ratio}). In conclusion we observe, therefore, that, for $r>3$, there is at most weak fragmentation if the total number of Krylov subspaces goes as $K\sim e^{c L}$. 
{\color{black} These findings are pictorially summarized by Fig.~\ref{fig: schema}.}



\section{Quantum chaos}
\begin{figure}
[h!]
\centering
\includegraphics[width=0.89\columnwidth]{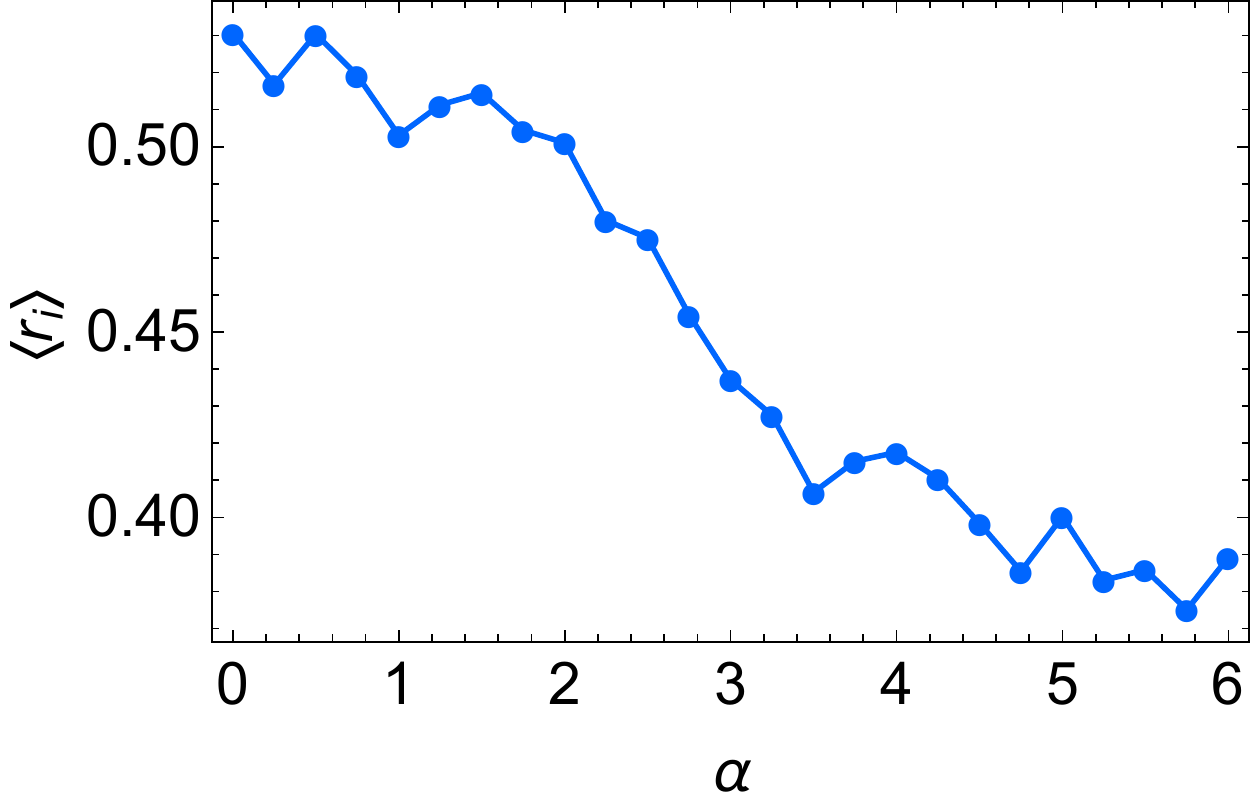}
\caption{$\langle r_i\rangle$ as a function of $\alpha$. We consider $r=3$, $L=16$ and the largest Krylov subspace in the sector with $S_z=0$, $S_1=1$ and $S_L=-1$.
}
\label{fig: qcr}
\end{figure}

Let us consider the energy level spacing statistics of the Krylov subspaces. We focus on the average energy level spacing ratio for the eigenenergies
in these subspaces, i.e., ${\sf r}_i=\min(\delta_i,\delta_{i+1})/\max(\delta_i,\delta_{i+1})$, where $\delta_i=E_i-E_{i+1}$ is the gap between adjacent energy levels~\cite{Oganesyan07}. The distribution of ${\sf r}_i$ has been studied in Ref.~\cite{Atas13}. If the energy spectrum has a Gaussian-orthogonal ensemble level statistics, then the average value is $\langle {\sf r}_i\rangle \approx 0.53$ for $L\to\infty$. In contrast, if the level statistics is Poissonian, then $\langle {\sf r}_i\rangle \approx 0.39$.
Differently from the fragmentation, the level spacing statistics depends on the specific values of the non-zero couplings $d_\ell$. We consider a power-law decay $d_\ell=1/\ell^\alpha$. For $r=3$, $L=16$, in the largest Krylov subspace of the sector defined by $S_z=0$, $S_1=1$ and $S_L=-1$,
we notice that, for large $\alpha$ (short range), $\langle {\sf r}_i\rangle$ is consistent with an integrable system, while decreasing $\alpha$, the system becomes more and more chaotic, and, for small $\alpha$ (long-range), $\langle {\sf r}_i\rangle$ is consistent with the Wigner-Dyson distribution (see Fig.~\ref{fig: qcr}).
\section{Scar states}
\begin{figure}
[t!]
\centering
\includegraphics[width=0.89\columnwidth]{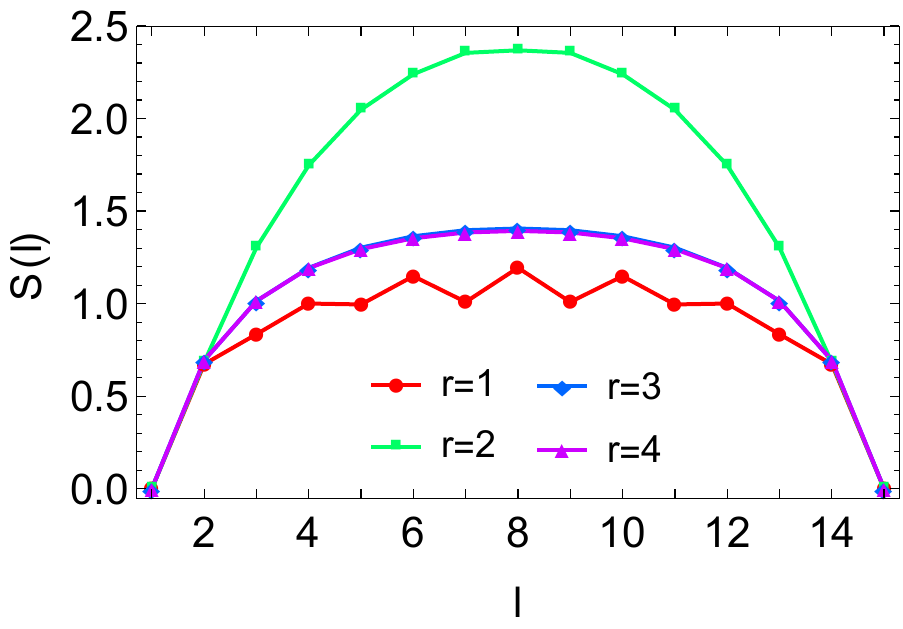}
\caption{Entanglement entropy $S(l)$ of a block of length $l$ for $L=16$ and different $r$. We consider the state {\color{black}$\ket{\Psi_n}$, Eq.~\eqref{eq. scars}, that is} in the largest Krylov subspace of the sector with $S_z=0$, $S_1=1$, $S_L=-1$ (for $r=1$ we have, in addition, $n_{\textrm{dw}}=-3$, and for $r=2$, $n_{\textrm{dw}}=1$).
}
\label{fig: scar}
\end{figure}

\begin{figure}
[t!]
\centering
\includegraphics[width=0.89\columnwidth]{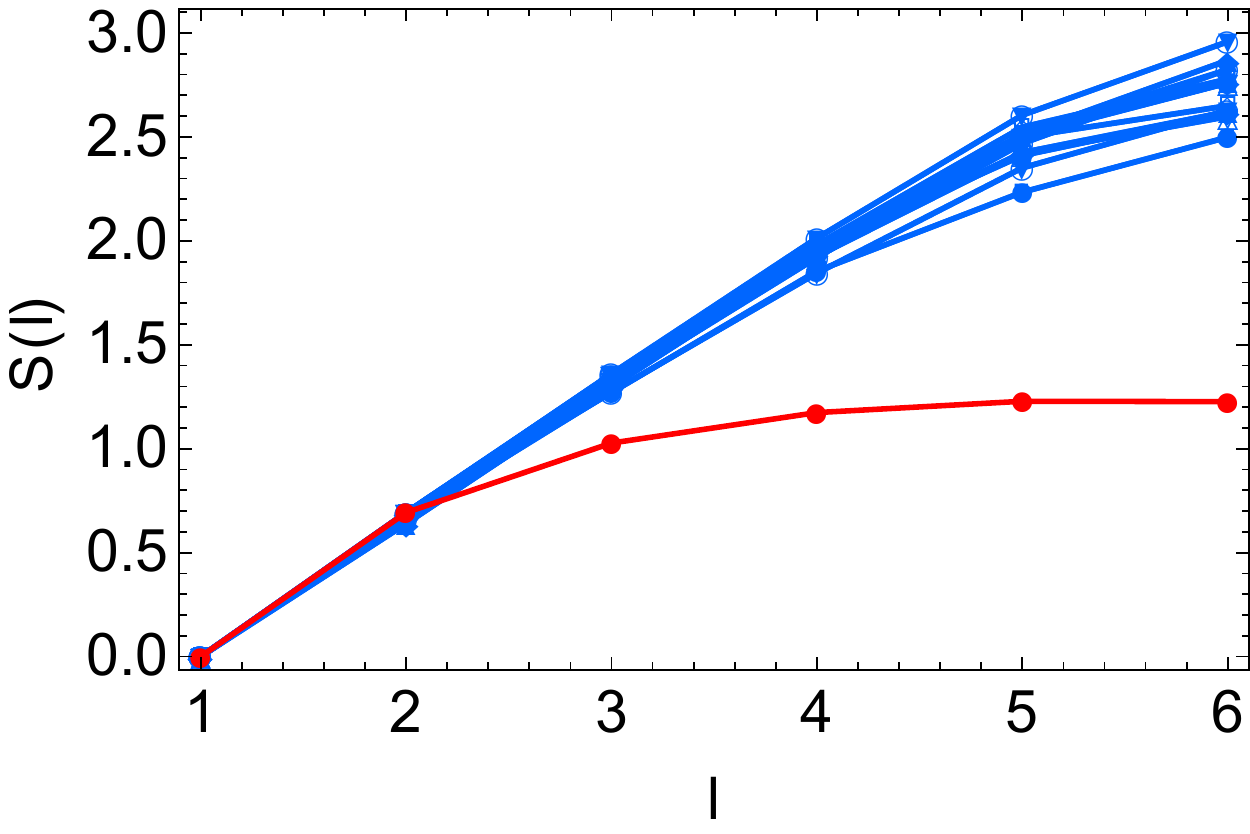}
\caption{{\color{black}Entanglement entropy $S(l)$ of a block of length $l$ for $L=12$, $r=3$ and $d_{\ell\leq r}=1$. We consider all the eigenstates in the largest Krylov subspace of the sector with $S_z=0$, $S_1=1$, $S_L=-1$, having energy in the middle of the spectrum. The red line corresponds to the scar state $\ket{\Psi_n}$. The blue lines correspond to the remaining states at zero energy, which show a thermal behavior, namely scales approximately linearly.}
}
\label{fig: scarVStherma}
\end{figure}

\begin{figure}
[t!]
\centering
\includegraphics[width=0.89\columnwidth]{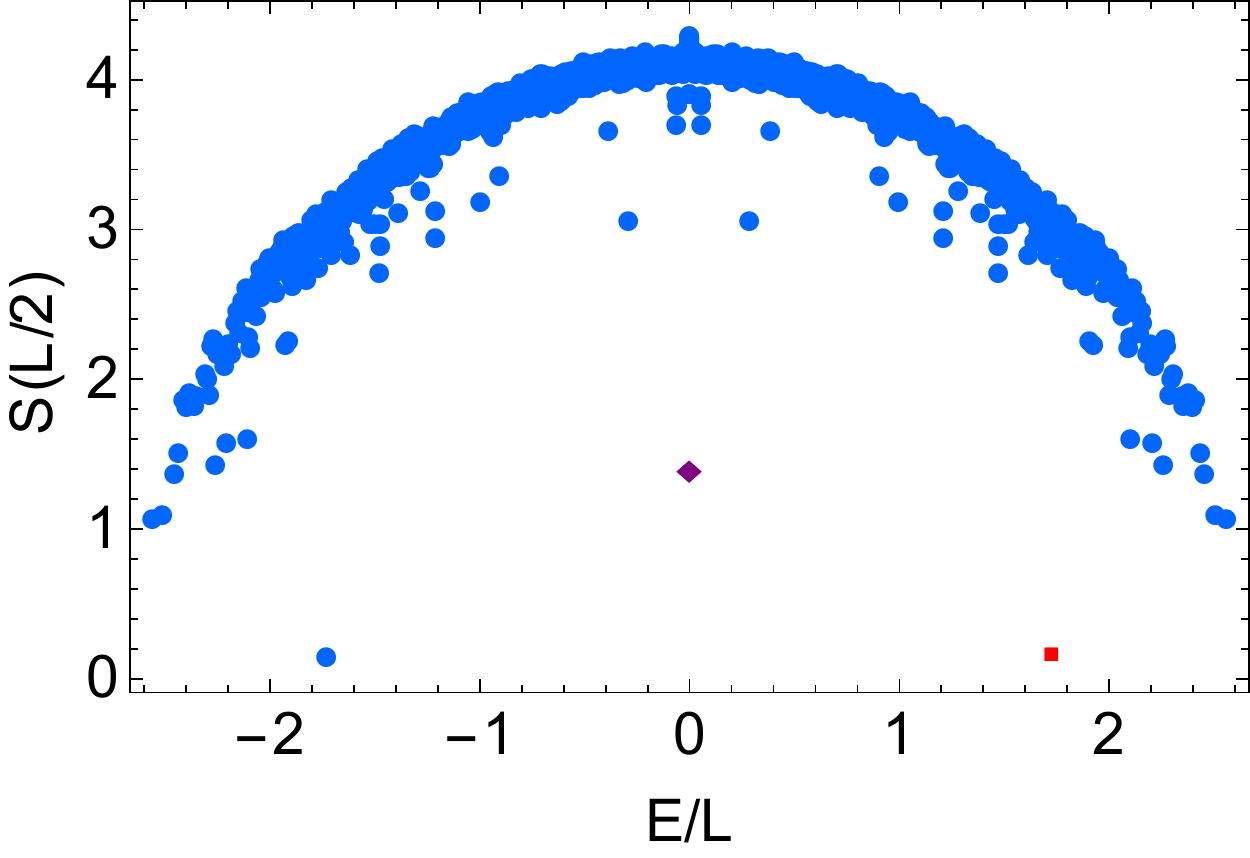}
\caption{Entanglement entropy $S(L/2)$ of all the energy eigenstates in the largest Krylov subspace in the sector with $S_z=0$, $S_1=1$ and $S_L=-1$, with $L=16$, $r=3$ and $d_{\ell\leq r}=1$. {\color{black}The diamond purple point corresponds to the scar state $\ket{\Psi_n}$, the  square red point corresponds to a scar state having a large overlap with the state $\ket{\psi_s}$ in Eq.~\eqref{psis}.}
}
\label{fig: ent ene}
\end{figure}

Given a Krylov subspace $\mathcal K_n$, we consider the state
\begin{equation}\label{eq. scars}
\ket{\Psi_n} = \mathcal N_n \sum_\alpha \ket{\psi^{(n)}_\alpha}\,,
\end{equation}
where $\ket{\psi^{(n)}_\alpha}$, {\color{black} with $\alpha=1,\dots,D_n$, are the states of the computational basis 
(namely, all the states of the form $\ket{s_1}\ket{s_2}\cdots\ket{s_L}$ with $s_i=\uparrow, \downarrow$)
which belong to} the Krylov subspace $\mathcal K_n$ with dimension $D_n$,
and $\mathcal N_n=1/\sqrt{D_n}$.
Noticing that $\otimes_{i=1}^L (\ket{\uparrow_i}+\ket{\downarrow_i})$ is equal to the sum of all the states in the computational basis, we get
\begin{equation}
\otimes_{i=1}^L (\ket{\uparrow_i}+\ket{\downarrow_i}) = \sum_n \frac{1}{\mathcal N_n} \ket{\Psi_n}\,.
\end{equation}
Moreover, $H\otimes_{i=1}^L (\ket{\uparrow_i}+\ket{\downarrow_i})=0$, then
\begin{equation}
\sum_n \frac{1}{\mathcal N_n} H\ket{\Psi_n}=0
\end{equation}
and since $H\ket{\Psi_n}\in \mathcal K_n$, all the states  $H\ket{\Psi_n}$ are linearly independent then $H\ket{\Psi_n}=0$ for every $n$. We calculate the entanglement entropy of $\ket{\Psi_n}$ for a block of length $l$, which is defined as $S(l)=-\Tr{\rho_{[1,l]} \ln \rho_{[1,l]}}$, where $\rho_{[1,l]}$ is the reduced density matrix of the block.
As shown in Fig.~\ref{fig: scar}, for the state $\ket{\Psi_n}$ in one of the largest Krylov subspaces, we find that, for $l<L/2$, the entropy $S(l)$ increases slower than a linear function of $l$.
In contrast, {\color{black} as shown in Fig.~\ref{fig: scarVStherma},} for the eigenstates with energies in the middle of the spectrum, the ETH predicts that the entropy is extensive (for energies in the middle of the spectrum the microcanonical ensemble has a very large temperature and from ETH we can give a random-matrix theory description, so that the entropy is given by the Page formula~\cite{Page93} restricted to the subspace and it is extensive).\\
Let us consider $r\geq 3$, which can host at most weak fragmentation. As shown in Fig.~\ref{fig: ent ene}, there are isolated eigenstates below the ETH-like curve, indicating the presence of further scar states. Considering one of those states,
the entanglement entropy $S(L/2)$ grows at most logarithmically with $L$, as shown in Appendix~\ref{app. scars}, therefore ETH is violated. 

\section{Out-of-equilibrium dynamics}
\begin{figure}
[t!]
\centering
\includegraphics[width=0.89\columnwidth]{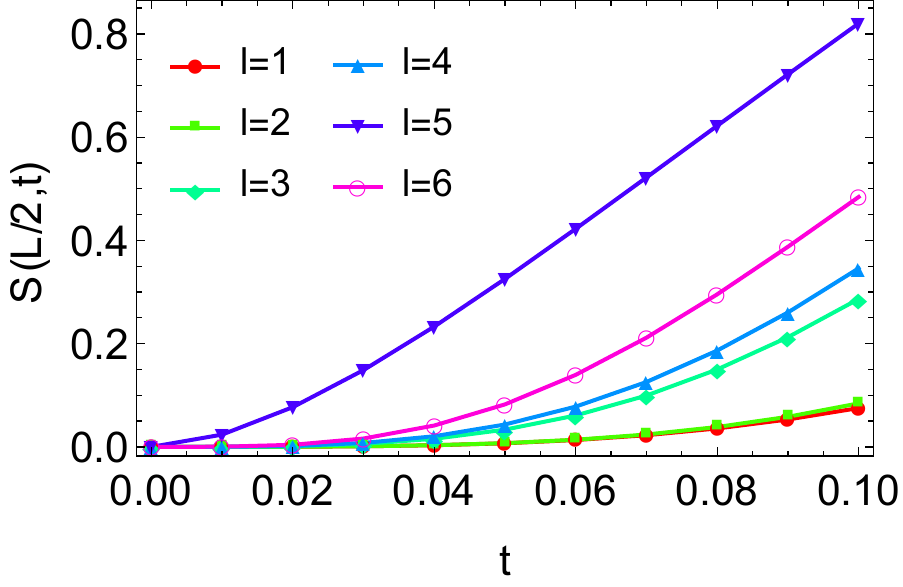}
\caption{Time-evolution of the entanglement entropy $S(L/2,t)$, with the initial states $\ket{\psi_l}$, for different values of $l$, and $r=3$, $L=16$.
}
\label{fig: entan time}
\end{figure}

In order to study the out-of-equilibrium dynamics we consider $r\geq 3$, with $d_{\ell}=1$ for any $\ell\leq r$, and even $L/2$. We will focus our attention to the largest Krylov subspace in the sector defined by $S_z=0$, $S_1=1$ and $S_L=-1$. Let us consider the following initial state
\begin{equation}
\label{psil}
\ket{\psi_l}= \ket{\uparrow}^{\otimes l} \otimes \ket{\uparrow \downarrow}^{\otimes L/2-l} \otimes \ket{\downarrow}^{\otimes l}\,,
\end{equation}
 which either belongs to the largest Krylov subspace or it is a frozen state, depending on the value of $l$. In detail, the initial state belongs to the largest Krylov subspace for $1\leq l \leq L/2-2$ while it is a frozen state for $l=0$ and $l=L/2-1$.
%
We calculate the time-evolution of the entanglement entropy of half-chain, $S(L/2,t)$,
starting from the initial state in Eq. (\ref{psil}).
As shown in Fig.~\ref{fig: entan time}, typically, initially, at early time, the entanglement entropy remains approximately null for a while before it starts growing.
In Ref.~\cite{Langlett21}, this behavior is explained by the long time needed for the spin up at site $l$ and the spin down at site $L-l+1$ to propagate towards the center of the chain before they scatter. However, in contrast to this behavior, we found that for the special case with 
$l=L/2-3$ the entropy starts growing immediately, see Fig. \ref{fig: entan time} for $r=3$ (we check that the same behavior occurs for $r\ge 3$). At later time the entropy grows linearly, therefore, the propagation of the excitations is ballistic.

\begin{figure*}[th!]
\centering
\begin{minipage}[b]{\textwidth}
\includegraphics[width=0.49\columnwidth]{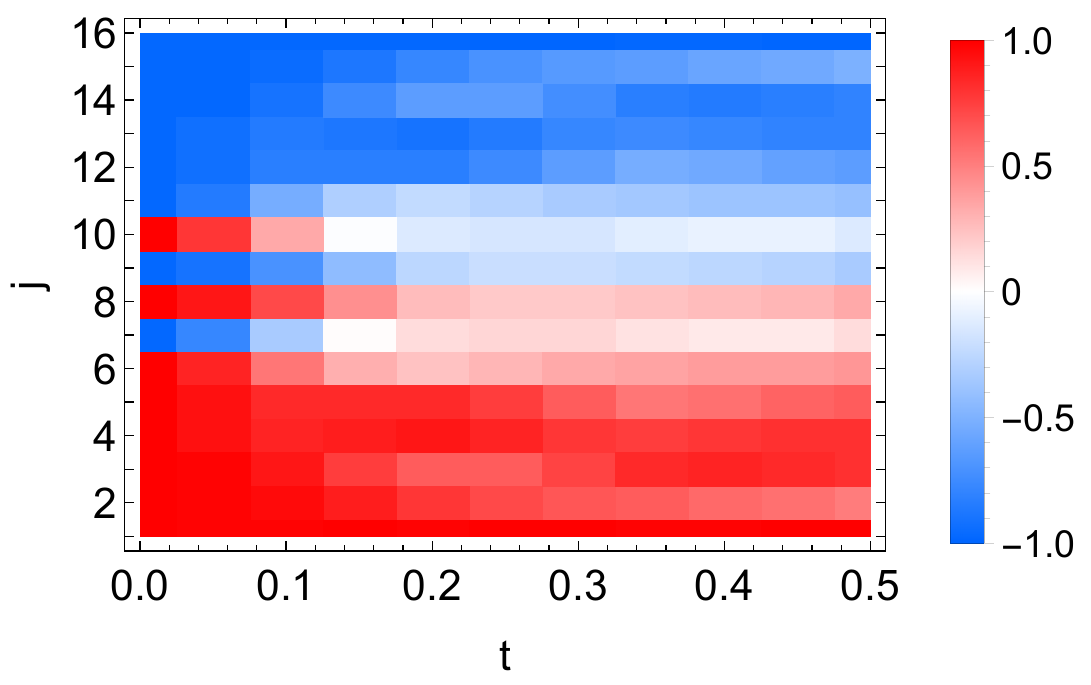}
\includegraphics[width=0.47\columnwidth]{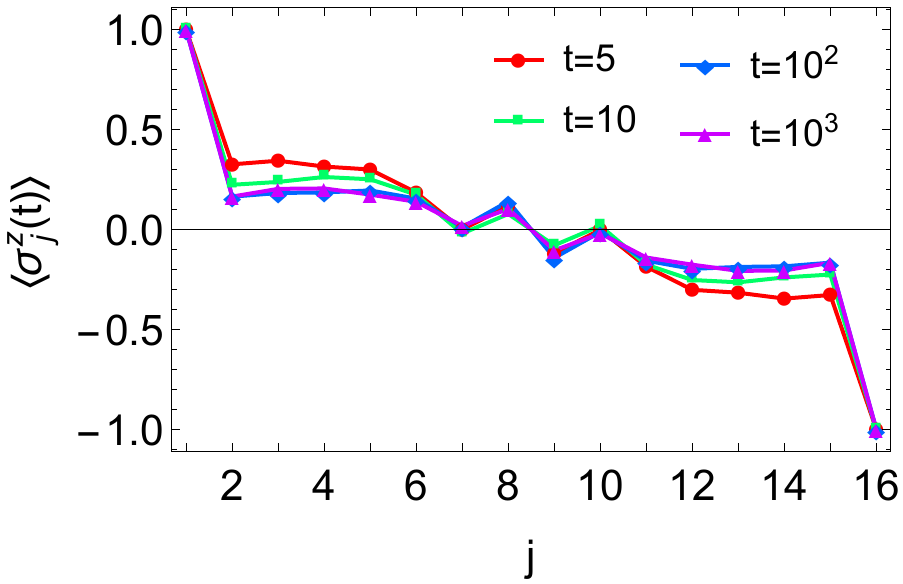}
\caption{(Left panel) Time-evolution of the magnetization $\langle \sigma_j^z(t) \rangle$, with the initial state $\ket{\psi_l}$, Eq. (\ref{psil}), with $l=5$ and $L=16$, for $r=3$. {\color{black}(Right panel) Magnetization profile at different long times.}
}
\label{fig: z time}
\end{minipage}%
\end{figure*}

The average value of the energy of $\ket{\psi_l}$ is $E=\bra{\psi_l}H\ket{\psi_l}=0$ which is in the middle of the spectrum, then from the ETH we expect that the time average of the reduced density matrix of a small part of the system tends to a thermal state with infinite temperature, so that the magnetization $\langle \sigma^z_j (t)\rangle$ tends to zero for $1<j<L$ at long times, while the boundary spins do not evolve. In Fig.~\ref{fig: z time} we study the time evolution of the magnetization $\langle \sigma_j^z(t) \rangle$.
\begin{figure*}[th!]
\centering
\begin{minipage}[b]{\textwidth}
\includegraphics[width=0.49\columnwidth]{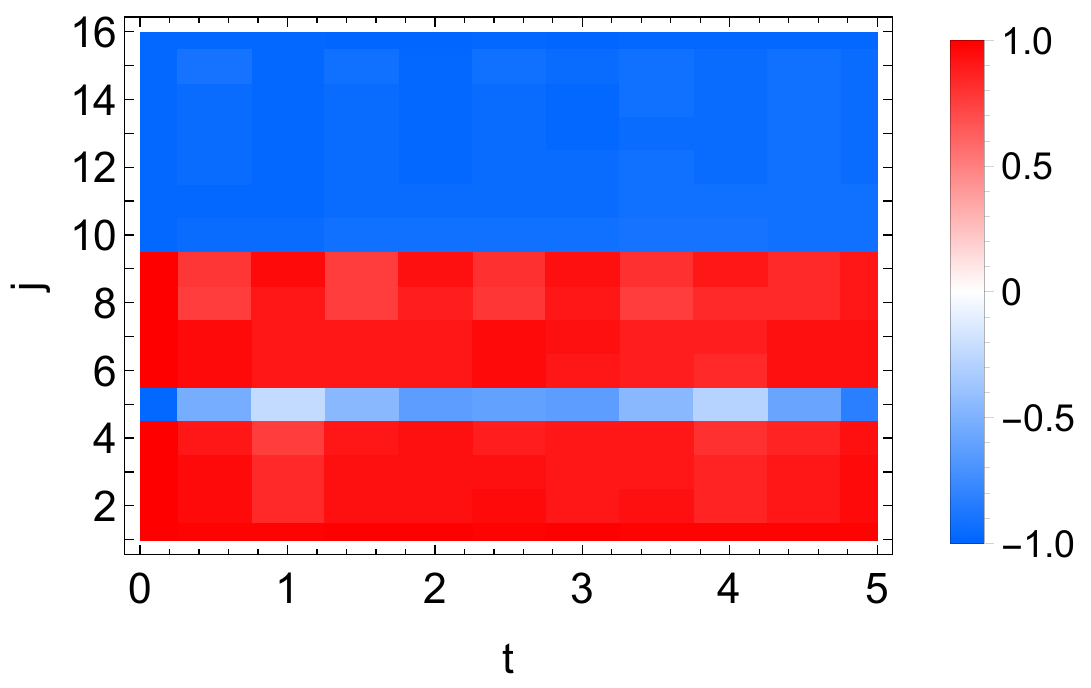} 
\includegraphics[width=0.47\columnwidth]{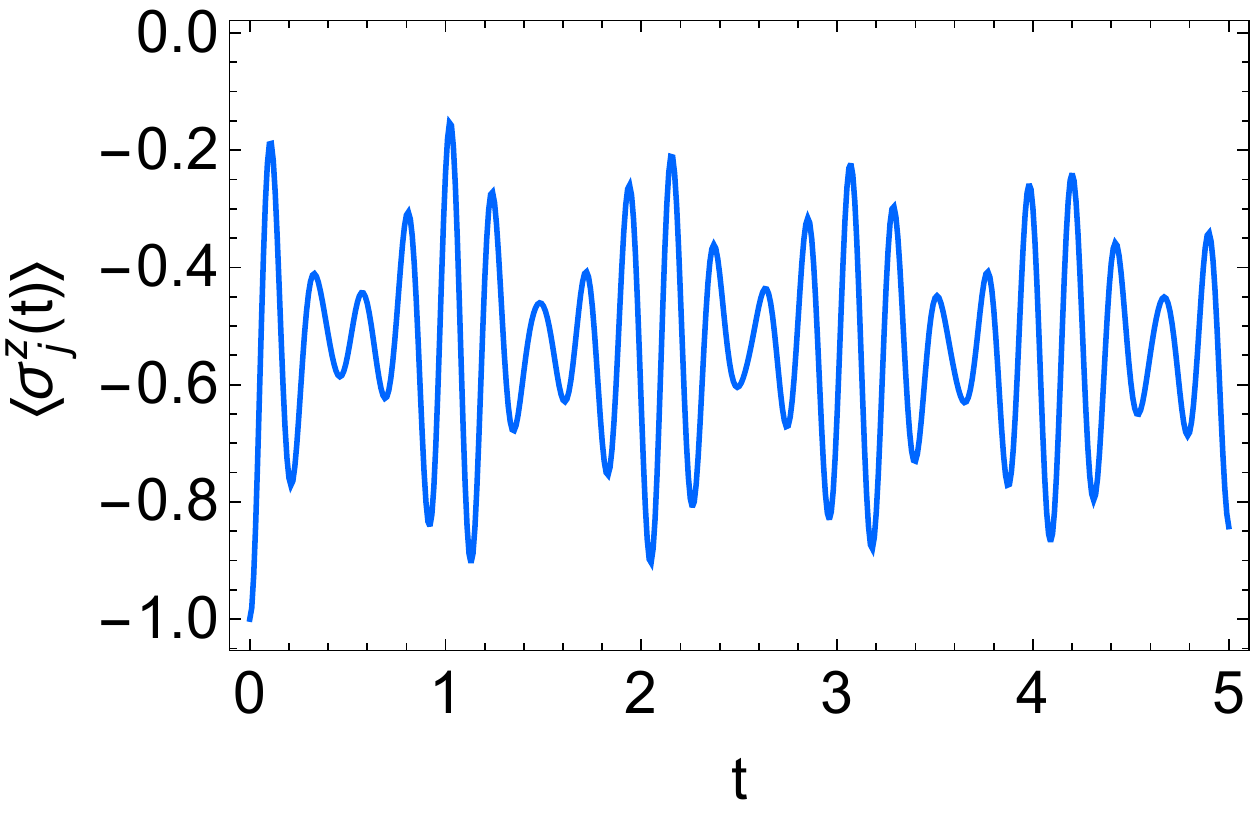}
\caption{(Left panel) Time-evolution of the magnetization $\langle \sigma_j^z(t) \rangle$ with the initial state $\ket{\psi_s}$, Eq. (\ref{psis}), for $L=16$ and $r=3$. {\color{black} (Right panel) Time-evolution of the magnetization $\langle \sigma_j^z(t) \rangle$ at the site $j=L/2-3$.}
}
\label{fig: z time scar}
\end{minipage}%
\end{figure*}
We note that at long times the magnetization tends to a stationary spatially non-uniform profile.
{\color{black}The weak fragmentation can contribute to this behavior}: although Krylov-restricted ETH is weakly satisfied in the dominant Krylov subspace, the initial state does not involve the eigenstates belonging to the non-dominant Krylov subspaces giving a non-uniform expectation value; this because the magnetization profile for the eigenstates in the non-dominant Krylov subspaces is non-uniform having predominantly an inversion of the spin directions (in the bulk) with respect to the stationary profile.
We note, indeed, that the magnetization profile for the completely mixed state restricted to the dominant Krylov subspace is $\langle \sigma^z_i\rangle = A_i/D_{max}$ in the bulk, where $A_i$ is predominantly positive (negative) in the left (right) part of the chain, whereas $\langle \sigma^z_i\rangle$ is zero in the bulk for the completely mixed state restricted to the full sector with definite symmetries.
{\color{black} However, since the fragmentation is weak, $A_i/D_{max}$ is small compared to the long-time magnetization value observed 
(since $A_i\sim (D-D_{max})\rightarrow 0$). Thus, we deduce that the scars will strongly contribute to the non-uniformity of the magnetization. In particular, the scar $\ket{\Psi_n}$ has also the magnetization $\langle \sigma^z_i\rangle = A_i/D_{max}$, so that the main contribution predominantly comes from other scar states.}
Moreover, due to the scar states we can have initial states which do not exhibit thermalization.
This behavior is very prominent, for $r=3$, using the following initial state
\begin{equation}
\label{psis}
\ket{\psi_s} =\ket{\uparrow}^{\otimes L/2-4}\otimes\ket{\downarrow}\otimes\ket{\uparrow}^{\otimes 4}\otimes\ket{\downarrow}^{\otimes L/2-1}\,,
\end{equation}
 which belongs to the largest Krylov subspace and has a large overlap with a high energy scar state.
 For $L=16$, the energy is $E=\bra{\psi_s}H\ket{\psi_s}=24$
 and the magnetization $\langle \sigma_j^z(t) \rangle$ stays almost equal to its initial value, except for the site $L/2-3$ at which the magnetization shows persistent oscillations, see Figs.~\ref{fig: z time scar}. 
 {\color{black} This behavior is due to the fact that the magnetization of the scar state differs from the state $\ket{\psi_s}$ only locally, near the site $L/2-3$.}
\section{Conclusions}
Fragmentation of the Hilbert space is believed to exist for interactions involving few sites. Here, we investigated the fate of fragmentation and scar states upon increasing the range of interaction by considering a modified long-range Fredkin spin chain which exhibits fragmentation due to interference.
We prove that there is fragmentation, i.e., there is an exponential number of Krylov subspaces in the thermodynamic limit, for any
range which is not asymptotically linear in the size $L$ with a linear coefficient equal to one.
{\color{black} The nearest neighbor case $r=1$ has been investigated also in Ref.~\cite{Langlett21}, where it is shown that there are scars and strong fragmentation with the standard exponential decay. Here, we find that for $r=1$ the strong fragmentation can also exhibit an algebraic decay of the ratio $D_{max}/D\sim 1/L$ in some sectors. By increasing the range, for a range $r=2$, we have both weak and strong fragmentation. Interestingly, in this case, the strong fragmentation exhibits only the algebraic decay of the ratio $D_{max}/D\sim 1/L$.}
For a range $r\ge 3$ there is only weak fragmentation.
Moreover, we show that scar states exist at any range of the interaction inside non-integrable Krylov subspaces. Finally, we investigate the effects of such structures of the Hilbert space on the out-of-equilibrium dynamics which is characterized by either local persistent oscillations or non-uniform stationary profile of the magnetization.
{\color{black} It might be interesting investigating the out-of-equilibrium dynamics through Krylov complexity as done in Ref. \cite{Nandy}.

\paragraph*{Note added:} After our work was completed, we became aware of the work \cite{Stephen} where an example of fragmentation, in two dimensions, robust to long-range perturbations is reported.

\subsection*{Acknowledgements}
The authors acknowledge financial support from the project BIRD 2021 "Correlations, dynamics and topology in long-range quantum systems" of the Department of Physics and Astronomy, University of Padova, 
and from the European Union-Next Generation EU within the "National Center for HPC, Big Data and Quantum Computing" (Project No. CN00000013, CN1 Spoke 10 - Quantum Computing).}

\appendix
\section{Calculation of the dimension of the largest Krylov subspace and the number of frozen states}\label{app. Krylov}
Given an operator $H$ and a state $\ket{\psi}$, the corresponding Krylov subspace is generated by the vectors $\ket{\psi}$, $H\ket{\psi}$, $H^2\ket{\psi}$, and so on. In detail, only the first $n$ vectors are linearly independent, for a certain integer $n$, which is the dimension of the Krylov subspace. A way to calculate $n$, is based on applying iteratively $H$ and using the Gram-Schmidt process at each step to  orthonormalize the set of vectors, so that, at the $(n+1)$-th step, one has a null vector. 
Let us consider a Krylov subspace $\mathcal K_n$ obtained from a root state $\ket{\psi_n}$.
{\color{black}
We consider all the product states $\ket{\psi^{(n)}_\alpha}$ in the computational basis 
generated by applying powers of $H$.
We have that $\ket{\psi^{(n)}_\alpha}\in \mathcal K_n$, then since $\ket{\psi^{(n)}_\alpha}$, for $\alpha=1,\ldots,D_n$, are $D_n$ linear independent states, we have that $D_n\leq \text{dim} \mathcal K_n$. In order to prove that $D_n = \text{dim} \mathcal K_n$, we can proceed ad absurdum. Let us assume that $D_n < \text{dim} \mathcal K_n$ for some $n$. We have that $\mathcal H = \bigoplus_n \mathcal K_n$ so that $\text{dim} \mathcal H =\sum_n \text{dim} \mathcal K_n$. Since all the states $\ket{\psi^{(n)}_\alpha}$ form the computational basis of $\mathcal H$, we get $\text{dim} \mathcal H = \sum_n D_n$, then $\text{dim} \mathcal H =\sum_n D_n < \sum_n \text{dim} \mathcal K_n$, which is a contradiction. This means that $\mathcal K_n$ is generated by the product states $\ket{\psi^{(n)}_\alpha}$ for $\alpha=1,\ldots,D_n$.
A simple method to determinate the dimension $D_{max}$ of the largest Krylov subspace is to apply $H$ to every product state belonging to the sector of dimension $D$.
On the other hand, to determinate the number $N_{froz}$ of Krylov subspaces of dimension one, called frozen states, we apply $H$ to all the product states of the full Hilbert space $\mathcal H$, and we count the number of those product states which are eigenstates of $H$.}

\section{Entropy of a scar state}\label{app. scars}
Let us show that the entanglement entropy of the scar state in Eq.~\eqref{eq. scars} grows at most logarithmically with $L$. For simplicity we consider
the following state, which, for large even $L$ and $S_z=0$ well approximates a scar state with $E=0$,
\begin{equation}
\ket{\Psi} = \frac{1}{\sqrt{N(L)}}\sum_{1\leq i_1<i_2< \cdots < i_{L/2}\leq L} \sigma^+_{i_1}\sigma^+_{i_2}\cdots \sigma^+_{i_{L/2}}\ket{\downarrow}^{\otimes L}\,,
\end{equation}
where $\sigma^+=\ket{\uparrow}\bra{\downarrow}$ and $N(L)=\binom{L}{L/2}$.  A string of local spins, like $\uparrow \uparrow \downarrow \ldots$ can be mapped to a lattice path by attaching a diagonal line $\diagup$ to $\uparrow$, and $\diagdown$ to $\downarrow$, from left to right. The state $\ket{\Psi}$ can be expressed as the sum over all the paths $w$ starting from $(0,0)$ and ending at $(L,0)$
\begin{equation}
\ket{\Psi} = \frac{1}{\sqrt{N(L)}} \sum_w \ket{w}\,.
\end{equation}
To calculate the reduced density matrix $\rho_{1,L/2}$, we consider the paths $w_y$ from $(0,0)$ to $(L/2,y)$ and the paths $z_y$ from $(L/2,y)$ to $(L,0)$, where $y=-L/2,-L/2+2,\ldots,L/2$, thus we can write
\begin{equation}
\ket{\Psi} = \frac{1}{\sqrt{N(L)}}\sum_y \sum_{w_y}\sum_{z_y} \ket{w_y}\otimes\ket{z_y}\,.
\end{equation}
We can calculate the partial trace by using the basis $\{\ket{z_y}\}$, thus we get
\begin{equation}
\rho_{[1,L/2]}= \frac{1}{N(L)}\sum_y N_y(L)\sum_{w_y}\sum_{w'_y} \ket{w_y}\bra{w'_y}\,,
\end{equation}
where $N_y(L)=\binom{L/2}{L/4+y/2}$ is the number of paths $w_y$. Then, the eigenvalues of $\rho_{[1,L/2]}$ are $\lambda_y = (N_y(L))^2/N(L)$, and the entanglement entropy is
\begin{equation}
S(L/2) = -\sum_y \lambda_y\ln \lambda_y \simeq \ln L\,.
\end{equation}
To derive this asymptotic formula we use the Stirling's approximation, so that
\begin{equation}
\lambda_y \simeq \frac{1}{\sqrt{L}}\frac{e^{L f(x_y)}}{1-x_y^2}\,,
\end{equation}
where $x_y=2y/L$ and $f(x)=-(1-x)\ln(1-x)-(1+x)\ln(1+x)$. For $L\to\infty$ we get
\begin{equation}
S(L/2) \simeq L \int_{-1}^1 \frac{1}{\sqrt{L}}\frac{e^{L f(x)}}{1-x^2} \ln\left(\frac{1}{\sqrt{L}}\frac{e^{L f(x)}}{1-x^2}\right) dx \,.
\end{equation}
Using the Laplace method, since $f(x)$ has only one maximum at $x=0$, we get $S(L/2) \simeq  \ln L\,$.

\end{document}